\documentclass[review]{elsarticle}

\usepackage{lineno,hyperref,amsmath,amssymb,xcolor}
\modulolinenumbers[5]
\journal{Wave Motion}

\newtheorem{theorem}{Theorem}
\newcommand{\id}{{1 \mskip -5mu {\rm I}}}

\newcommand{\rr}[1]{{\normalfont\textrm{#1}}}

\newcommand{\bb}[1]{{\mathbb{#1}}}
\newcommand{\mc}[1]{{\mathcal #1}}

\begin{document}

\begin{frontmatter}

\title{Stability of the stationary solutions of 
the Allen--Cahn equation with non--constant stiffness}

\author[addP]{Paolo Butt\`a}
\ead{butta@mat.uniroma1.it}

\author[addE]{Emilio N.M.\ Cirillo}
\ead{emilio.cirillo@uniroma1.it}

\author[addG]{Giulio Sciarra}
\ead{giulio.sciarra@ec-nantes.fr}



\address[addP]{Dipartimento di Matematica,
	Sapienza Universit\`a di Roma, 
	piazzale Aldo Moro 5, I--00185, Roma, Italy.}

\address[addE]{Dipartimento di Scienze di Base e Applicate per 
	l'Ingegneria, Sapienza Universit\`a di Roma, 
	via A.\ Scarpa 16, I--00161, Roma, Italy.}

\address[addG]{Institut de Recherche en G\'enie Civil et M\'ecanique,
             Ecole Centrale de Nantes,
             1, rue de la No\"{e} 44321 Nantes, France}

\begin{abstract}
We study the solutions of a generalized Allen--Cahn equation
deduced from a Landau energy functional, endowed with a non--constant
higher order stiffness.
We assume the stiffness to be a positive
function of the field and
we discuss the stability of the
stationary solutions proving both linear and local non--linear stability.
\end{abstract}


\end{frontmatter}

\linenumbers

\section{Introduction}
\label{s:introduzione} 
The behavior of homogeneous phases simoultaneously 
present in a physical 
system can be described via a \emph{phase--field} 
$\phi(x,t)$ depending on the space variable
$x\in\Omega\subset\bb{R}^3$ 
and on the time variable 
$t\in[0,\infty)$.
Two particular values of the field 
represent the two homogeneous phases.
These models have been widely used in the study 
of the \emph{spinodal decomposition} phenomenon \cite{Bray,Langer,Eyre}, 
namely, the process in which a system undergoing a 
second order phase transition is suddenly 
quenched from the disordered high temperature phase into a 
broken--symmetry low temperature state
and the evolving field $\phi(x,t)$ encodes 
the process of separation of the low temperature phases
\cite{CCMMMS}.

The evolution equation can be obtained as 
a gradient equation \cite{Fife,ABF}
of a suitable Landau energy functional, namely,
\begin{equation}
\label{grad}
\frac{\partial \phi}{\partial t} = -\rr{grad}\,H(\phi)
\;\;\;\textup{ with }\;\;\;
H(\phi)
:=
\int_\Omega\Big[\frac{1}{2}\varepsilon|\nabla \phi|^2
+ W(\phi)\Big]\,\rr{d}x
\,,
\end{equation}
where the symbol grad means the variational derivative of the functional $H$.
The function
$W\in C^2(\bb{R})$ is the bulk energy of the field $\phi$ and 
the \textit{higher--order stiffness} (or, simply, stiffness) 
$\varepsilon$
weights the interface contribution represented by the
squared--gradient term.
The bulk energy $W$ is usually chosen as a 
a double well with the two minima 
corresponding to the two phases $0$ and $1$.
If the stiffness is constant and 
no constraint to the total value of the field $\phi$ is imposed, 
the \textit{standard} Allen--Cahn or Ginzburg--Landau equation 
\begin{equation}
\label{ac00}
\frac{\partial \phi}{\partial t}
=
\varepsilon\Delta \phi-W'(\phi)
\end{equation}
is found, see, e.g., \cite{Allen}, which was firstly introduced to 
describe the motion of anti--phase boundaries in crystalline solids. 
Recently a similar model has been introduced also in the framework 
of the mechanics of (partially) saturated porous media, 
see, e.g., \cite{CIS2010,CIS2013, Sciarra2016}.

In this paper we consider the case in which the higher--order stiffness 
is not constant \cite{BSBS,FJ,CIS2015}, but it is a sufficiently regular 
positive function of the field, namely, $\varepsilon\in C^\infty(\bb{R})$ 
such that there is $\bar C>1$ for which 
\begin{equation}
\label{csigma}
\frac{1}{\bar C} \le \varepsilon(\phi) \le \bar C\,.
\end{equation}
The fact that $\varepsilon$ is strictly positive will be crucial in 
this paper, indeed, it has been proven in different contexts, 
see, e.g., \cite{CIS2015,CCS2019,DGS2007},
that if the stiffness is allowed to vanish  
in correspondence of the homogeneous states, not regular 
stationary
solutions, often called \emph{compactons} \cite{RZ2018}, can be found. 

We also assume that the double well function $W\in C^\infty(\bb R)$ 
and satisfies, for some $\bar r\in (0,1)$,
\begin{equation}
\label{condW}
\begin{split}
& W(0) = W(1) = 0\,, \\ & W'(r) > 0 \text{ for } r\in (0,\bar r)\,, \\ 
& W'(r) <0 \text{ for } r\in (\bar r,1)\,, \\ & W''(0)>0\,, \;\; W''(1)>0\,.
\end{split}
\end{equation}

In the not constant stiffness case, 
the gradient equation \eqref{grad} provides the 
generalized Allen--Cahn equation 
\begin{equation}
\label{ac010}
\frac{\partial \phi}{\partial t} = \frac{1}{2}\varepsilon'(\phi)
|\nabla \phi|^2 + \varepsilon(\phi) \Delta \phi  -W'(\phi)
\end{equation}
see, for instance, \cite[Appendix~A]{CIS2015}.

It is useful to reformulate the problem in terms of the field 
\[
u(x,t) := g(\phi(x,t))\,, \qquad g(\phi):=\int_{1/2}^\phi\! \sqrt{\varepsilon(s)}\,\rr{d}s\,.
\]
Indeed, a straightforward computation shows that the Landau energy functional \eqref{grad} and the evolution equation \eqref{ac010} expressed in terms of the new field $u$ read, 
\[
F(u) = \int_\Omega\Big[\frac{1}{2}|\nabla u|^2+\tilde W(u)\Big]\,\rr{d}x
\]
and 
\begin{equation}
\label{ac010p}
\frac{\partial u}{\partial t} = - \sigma(u) \,\rr{grad}\,F(u) =  \sigma(u) \big[\Delta u  - \tilde W'(u) \big]\,,
\end{equation}
where $\tilde W := W \circ g^{-1}$ 
is still double well function and 
$\sigma := \varepsilon \circ g^{-1}$ plays the role of a \textit{mobility} for the dynamics. 


In the sequel, we shall consider the problem formulated in terms of the field $u$. Moreover, via an affine transformation of $u$, we can assume that the double well function, again denoted by $W$, still assumes its minima in $u=0,1$ and equations \eqref{condW} hold (with a different $\bar r$).

We focus on the one--dimensional case $\Omega=\bb{R}$ and study the evolution problem 
\begin{equation}
\label{ac020}
\left\{
\begin{array}{l}
{\displaystyle
\frac{\partial u}{\partial t} = \sigma(u) \Big[\frac{\partial^2u}{\partial x^2} -W'(u)\Big] }
\vphantom{\Big\{_\}}
\\
u(x,t) \stackrel{x\to-\infty}{\longrightarrow}0
\;\;\textrm{ and }\;\;
u(x,t)
\stackrel{x\to+\infty}{\longrightarrow}1
\\
\end{array}
\right.
\end{equation}
namely, we solve the equation with Dirichlet boundary conditions $0$ and $1$ respectively at minus and plus infinity. As $\sigma>0$, the associated stationary problem reads,
\begin{equation}
\label{ac040}
\left\{
\begin{array}{l}
{\displaystyle
u''(x)-W'(u(x)) =0
}
\vphantom{\Big\{_\}}
\\
u(x)
\stackrel{x\to-\infty}{\longrightarrow}0
\;\;\textrm{ and }\;\;
u(x)
\stackrel{x\to+\infty}{\longrightarrow}1
\\
\end{array}
\right.
\end{equation}
and is often 
addressed to in the literature as the problem of finding a 
\emph{connection} between the two phases on 
the line \cite{AF,Gurtin}.  

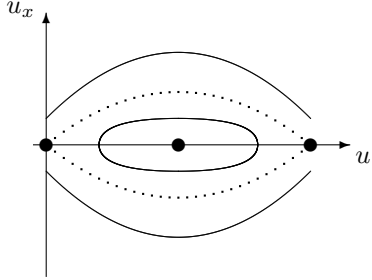
\begin{figure}[t]
\begin{picture}(200,130)(-120,-10)
\thinlines
\put(0,0){\vector(0,1){100}}
\put(-5,50){\vector(1,0){120}}
\put(0,50){\circle*{5}}
\put(50,50){\circle*{5}}
\put(100,50){\circle*{5}}
\thicklines
\qbezier[30](0,50)(50,90)(100,50)
\qbezier[30](0,50)(50,10)(100,50)
\thinlines
\qbezier(0,60)(50,110)(100,60)
\qbezier(0,40)(50,-10)(100,40)

\qbezier(20,50)(20,60)(50,60)
\qbezier(50,60)(80,60)(80,50)
\qbezier(20,50)(20,40)(50,40)
\qbezier(50,40)(80,40)(80,50)

\put(117,43){${u}$}
\put(-15,100){${u_x}$}
\end{picture}
\caption{Phase portrait of the 
stationary equation \eqref{ac040}.
The dotted lines represent the separatrix.
}
\label{f:qual}
\end{figure}

The existence of the solution to the connection problem is an immediate consequence of a simple qualitative analysis of the equation \eqref{ac040} based on the conservation law,
\begin{equation}
\label{ac050}
\frac{\rr{d}}{\rr{d}x}
\Big[
\frac{1}{2}(u')^2-W(u)
\Big]
=0 \,.
\end{equation}
It is easily seen that there exists a unique solution up to space translations, precisely the function $\bar u(x)$ defined by
\begin{equation}
\label{pao1}
x = \int_{1/2}^{\bar u(x)}\! \sqrt{\frac{1}{2W(s)}}\, \rr{d}s\,, \quad x\in\bb R\,,
\end{equation} 
together with its translates $\bar u_z(x) = \bar u(x-z)$. 
The phase portrait of the solutions to \eqref{ac050} in the 
interval $u\in[0,1]$ is plotted in figure~\ref{f:qual}. 
The separatrix is represented by a dotted line and its portion 
in the half--plane $u'>0$ corresponds to the solution of the 
connection problem. 

In this paper we discuss the stability of such stationary solutions,
hereafter named standing waves or \textit{fronts}, 
following ideas developed in \cite{FM} in the case of the 
standard Allen--Cahan evolution, see, also, 
\cite{BBB}.
We prove both
linear stability in Section~\ref{s:lineare}
and local non--linear stability in Section~\ref{s:nonlineare}. 
Some numerical simulations will be finally discussed in the closing 
Section~\ref{s:num}.

\section{Linear stability}
\label{s:lineare}
We first derive the linearized problem associated to \eqref{ac040}. We consider the case of the reference solution $\bar u$, the adaptation to the case of its translates $\bar u^z$ is immediate. 

We first emphasize some relevant properties of the connection $\bar u$. 
In view of the assumptions on $W$ and $\sigma$, by \eqref{pao1} it follows that $\bar u\in C^\infty(\bb R)$ with $\bar u'(x) >0$ for any $x\in \bb R$. Moreover, $\bar u$ approaches its limit exponentially fast as $x\to\pm\infty$; more precisely, there exist $A_0,A_1>0$ such that
\begin{equation}
\label{pao2}
\lim_{x\to-\infty}\rr{e}^{-\alpha_0 x}\bar u(x) = A_0\,, \; \lim_{x\to+\infty}\rr{e}^{\alpha_1 x}(1-\bar u(x)) = A_1\,,
\end{equation}
where $\alpha_0 = \sqrt{W''(0)}$ and $\alpha_1 = \sqrt{W''(1)}$. 

Finally, as $\bar u' = \sqrt{2W(\bar u)}$, by \eqref{pao2} a similar estimate holds for the derivative: 
\begin{equation}
\label{pao3}
\lim_{x\to-\infty}\rr{e}^{-\alpha_0 x}\bar u'(x) = A_0'\,, \;\; \lim_{x\to+\infty}\rr{e}^{\alpha_1 x}\bar u'(x) = A_1'\,,
\end{equation}
with $A_0'=A_0\sqrt{W''(0)}$ and $A_1'=A_1\sqrt{W''(1)}$.

Letting $u=\bar u+v$ and using that $\bar u$ is a stationary solution, to the first order in $v$ the problem \eqref{ac020} yields,
\begin{equation}
\label{li010}
\left\{
\begin{array}{l}
{\displaystyle
\frac{\partial v}{\partial t}
=
\sigma(\bar u)\Big[\frac{\partial^2v}{\partial x^2} - W''(\bar u)v\Big]
}
\vphantom{\Big\{_\}}
\\
v(x,t)
\stackrel{x\to-\infty}{\longrightarrow}0
\;\;\textrm{ and }\;\;
v(x,t)
\stackrel{x\to+\infty}{\longrightarrow}0
\\
\end{array}
\right.
\end{equation}

By using again that $\bar u$ is a solution of the stationary problem, it is possible to cancel the term depending on the potential energy $W$ in the linearized problem. Indeed, by differentiating the stationary equation \eqref{ac040} we have 
\begin{equation}
\label{li020}
\bar u'''(x) = W''(\bar u(x))\bar u'(x)\,.
\end{equation}
By substituting this expression in the equation \eqref{li010} we have that the linearized equation can be written as 
\begin{equation}
\label{li030}
\frac{\partial v}{\partial t}
=
Lv
\end{equation}
with $L$ the linear operator 
\begin{equation}
\label{li040}
L = a(x)\Big[\frac{\rr{d}^2}{\rr{d}x^2} - \frac{\bar u'''(x)}{\bar u'(x)}\Big]\,, \qquad a(x) := \sigma(\bar u(x))\,,
\end{equation} 
on the space $L^2(\bb R;\rr{d}x)\cap C^2(\bb R)$ or $C_b(\bb R)\cap C^2(\bb R)$, where $C_b(\bb R)$ is the space of bounded continuous functions. We remark that, by the assumptions on $\sigma$, the function $a(x)$ is smooth, bounded, and such that $\inf_x a(x) >0$.

By \eqref{li020} it follows that $\bar u'$ is an eigenfunction of $L$ with eigenvalue zero. Our task is to show that the rest of the spectrum is strictly negative. To this end, we take advantage of the fact that $\bar u'>0$ to reduce the problem to estimating the rate of convergence to equilibrium of a suitable diffusion process.

We notice that $L\bar u' = \bar u' \mc L$, where
\begin{equation}
\label{pao4}
\mc L = a(x) \Big[\frac{\rr{d}^2}{\rr{d}x^2} - U'(x)\frac{\rr{d}}{\rr{d}x}\Big]\,,
\end{equation}
with 
\begin{equation}
\label{pao5}
U(x) = - 2\log\bar u'(x)\,.
\end{equation}
The operator $\mc L$ is the generator of a diffusion. More precisely, it is the generator of the Markov semigroup associated to the dynamics on $\bb R$ given by the solution to the stochastic differential equation,
\[
\rr{d} x = a(x) U'(x)\,\rr{d} t + \sqrt{2a(x)}\, \rr{d}B\,,
\]
where $B=B(t)$ is a 1-dimensional Wiener process. Since $a>0$, by elliptic regularity the transition probability $P_t(x,\rr{d}y)$ admits a smooth density $p_t(x,y)$ which is the fundamental solution associated to $\mc L$, i.e., 
\begin{equation}
\label{paof}
\begin{cases}
{\displaystyle \frac{\partial p_t}{\partial t}(x,\cdot) = \mc L p_t(x,\cdot)} \\ p_0(x,y) = \delta_y(x)
\end{cases}
\end{equation}
Moreover, $p_t(x,y)>0$ for any $x,y\in\bb R$ and $t>0$, hence the associated Markov process is irreducible aperiodic. 

Under these conditions, the process has a unique (up to multiplication with a constant) invariant measure with a smooth positive density \cite{Rey-Bellet}. This density is solution to the stationary Fokker-Plank equation $\mc L^*\varrho=0$, where $\mc L^*$ is the $L^2$-adjoint of $\mc L$. An explicit computation shows that $\varrho(x) = \bar u'(x)^2a(x)^{-1} = \rr{e}^{-U(x)}a(x)^{-1}$. By \eqref{pao3}, $\varrho$ is integrable so that we obtain an invariant probability measure,
\[
\mu(\rr{d}x) = \frac {\bar u'(x)^2a(x)^{-1}\,\rr{d}x}{\int_{\bb R}\! \bar u'(x)^2a(x)^{-1} \,\rr{d}x'}\,.
\]
It is worthwhile to notice that the process is in fact reversible with respect to $\mu$, since by integration by parts,
\[
\int\! \psi \mc L \varphi \, \rr{d}\mu = \int\! (\mc L \psi) \varphi\, \rr{d}\mu = - \int\! \sigma(\bar u) \psi'\varphi' \, \rr{d}\mu \,.
\]

To study the convergence $P_t \to \mu$ as $t\to \infty$ we apply a classical result based on the existence of a Liapunov function. By definition, a Liapunov function is a function $V\ge 1$ with compact level sets (i.e., $V(x) \to +\infty$ as $|x|\to \infty$). We claim there exists a Liapunov function $V$ for which there are a compact set $K\subset \bb R$ and constants $r>0$ and $b<\infty$ such that 
\begin{equation}
\label{pao6}
\mc L V(x)  \le -r V(x) + b \id_K(x)\,.
\end{equation}
Then, see, e.g., \cite[Thm.\ 8.7]{Rey-Bellet}, introducing the weighted norms,
\[
\|\varphi\|_{V,\infty} = \sup_{x\in \bb R} \frac{|\varphi(x)|}{V(x)}\,, \quad \|\varphi\|_{V,p} = \Big\|\frac{\varphi}{V}\Big\|_{ L^p(\bb R;\rr{d}x)}
\]
and letting 
\[
P_t\varphi(x) = \int\! p_t(x,y) \varphi(y) \, \rr{d}y \,, \quad \mu(\varphi) = \int\!\varphi \,\rr{d}\mu\,,
\]
there are constants $C>0$ and $\alpha>0$ such that the bounds
\begin{equation}
\label{pao7}
\begin{split}
\|P_t\varphi - \mu(\varphi)\|_V & \le C \rr{e}^{-\alpha t} \|\varphi-\mu(\varphi)\|_V\,, \\ 
\|P_t\varphi - \mu(\varphi)\|_{V,p} & \le C \rr{e}^{-\alpha t} \|\varphi-\mu(\varphi)\|_{V,p}
\end{split}
\end{equation}
hold for every measurable function $\varphi$ with finite weighted norm. 

We now prove the claim on the existence of $V$ for which \eqref{pao6} holds. By \eqref{pao3}, the function $V=\gamma\rr{e}^{U/2} = \gamma(\bar u')^{-1}$, with $\gamma = \max\bar u'$, is Liapunov and, by an explicit computation,  
\begin{equation}
\label{pao8}
\mc L V = \frac a2 \Big(U''-\frac 12 (U')^2\Big)\, V \,.
\end{equation}
Now, by \eqref{ac040}, \eqref{li020}, in view of \eqref{pao2}, and \eqref{pao3},
\[
\lim_{x\to-\infty} \frac{\bar u''(x)}{\bar u'(x)} = \sqrt{W''(0)}\,, \quad \lim_{x\to+\infty} \frac{\bar u''(x)}{\bar u'(x)} = \sqrt{W''(1)}\,,
\]
\[
\lim_{x\to-\infty} \frac{\bar u'''(x)}{\bar u'(x)} = W''(0)\,, \quad \lim_{x\to+\infty} \frac{\bar u'''(x)}{\bar u'(x)} = W''(1)\,.
\]
By computing $U'$ and $U''$ and using the above limits, a straightforward computation shows,
\[
\begin{split}
& \lim_{x\to-\infty} \frac{a(x)}2 \Big(U''(x)-\frac 12 U'(x)^2\Big) = - \sigma(0)  W''(0)\,, \\ & \lim_{x\to+\infty}  \frac{a(x)}2 \Big(U''(x)-\frac 12 U'(x)^2\Big) = -\sigma(1)  W''(1)\,.
\end{split}
\]
Therefore, in view of \eqref{pao8}, the function $V$ clearly satisfies \eqref{pao6}, e.g., with $r = \frac12 W''(0)\wedge W''(1) $ and $K=[-R,R]$ so large that $V>4$ and $\frac 12 a \Big(U''-\frac 12 (U')^2\Big) > r$ outside $K$.

The desired spectral properties of the operator $L$ can be now easily deduced from \eqref{pao7}. To this end, it is useful to work in the Hilbert space $L^2(\bb R;a(x)^{-1}\rr{d}x)$, where $L$ is symmetric. We denote the inner product by $\langle\cdot,\cdot\rangle_a$ and the corresponding norm by $\|\cdot\|_{a,L^2}$. We observe that since $V=\gamma(\bar u')^{-1}$, for every $v\in C_b(\bb R)$ or $v\in L^2(\bb R;\rr{d}x)$ the function $\varphi_v := v/\bar u'$ has the corresponding weighted norms finite. Moreover, as $\rr{e}^{Lt} = \bar u' P_t (\bar u')^{-1}$,
\[
\rr{e}^{Lt} v - \langle v, q \rangle_a q = \big[P_t\varphi_v - \mu(\varphi_v)\big]\bar u'\,,
\]
where $q(x) = \|\bar u'\|_{a,L^2}^{-1}\bar u'(x)$ is the normalized eigenfunction of $L$ with eigenvalue zero. Therefore, by \eqref{pao7}, it is easily deduced that there is a constant $C_0>0$ such that 
\begin{equation}
\label{pao7bis}
\begin{split}
\|\rr{e}^{Lt} v - \langle v, q \rangle_a q\|_{L^2} & \le C_0 \rr{e}^{-\alpha t} \|v-\langle v, q \rangle_a q\|_{L^2}\,, \\ 
\|\rr{e}^{Lt} v - \langle v, q \rangle_a q\|_\infty & \le C_0 \rr{e}^{-\alpha t} \|v-\langle v, q \rangle_a q\|_\infty\,,
\end{split}
\end{equation}
where, in the first estimate, we used that, in view of the assumptions on $\sigma$, the norms $\|\cdot\|_{a,L^2}$ and $\|\cdot\|_{L^2}$ are equivalent. 

An estimate like \eqref{pao7bis} holds also with respect to the $H^1$-norm. This is quite standard but we sketch the proof for the sake of completeness. To this purpose, we shall use the following estimates on the fundamental solution $\rr{e}^{Lt}(x,y)$, which can be deduced by exploiting the parametric method to construct the fundamental solution, see \cite{F}. Letting 
\begin{equation}
\label{paop0}
\rr{e}^{Lt}(x,y) = \sqrt{\frac{a(y)}{4\pi t}}\exp\big(-\frac{a(y)(x-y)^2}{4t}\big) + Q_t(x,y)\,,
\end{equation}
for any $T>0$ and $\kappa>\sup_x a(x)$ there exists a constant $C_T>0$ such that, for any $t\in (0,T]$ and $x,y\in\bb R$,
\begin{equation}
\label{paop}
\begin{split}
Q_t(x,y) & \le C_T\exp\Big(-\kappa\frac{(x-y)^2}{4t}\Big)\,, \\ 
\Big|\frac{\partial Q_t(x,y)}{\partial x}\Big| & \le \frac{C_T}{\sqrt t} \exp\Big(-\kappa\frac{(x-y)^2}{4t}\Big)\,.
\end{split}
\end{equation}

Let now $v\in H^1$ that, without loss of generality, we identify with its continuous representative. Clearly $\rr{e}^{Lt} v - \langle v, q \rangle_a q = \rr{e}^{Lt} v_\perp$, where $v_\perp = v - \langle v, q \rangle_a q$ is the component of $v$ orthogonal to $q$ in $L^2(\bb R;a(x)^{-1}\rr{d}x)$. We estimate $\big\|\rr{e}^{Lt} v_\perp\big\|_{H^1}$ by analyzing separately the case $t\in (0,1]$ and $t>1$. In view of \eqref{pao7bis}, we only need to study the $L^2$-norm of $\frac{\partial(\rr{e}^{Lt} v_\perp)}{\partial x}$. 

If $t\in (0,1]$, in view of the identity
\[
\begin{split}
& \frac{\partial}{\partial x} \frac{\rr{e}^{-\frac{a(y)(x-y)^2}{4 t}}}{\sqrt{4\pi a(y)t}} + \frac{\partial}{\partial y} \frac{\rr{e}^{-\frac{a(y)(x-y)^2}{4 t}}}{\sqrt{4\pi a(y)t}} = \frac{a'(y) \rr{e}^{-\frac{a(y)(x-y)^2}{4 t}}}{a(y)\sqrt{4\pi a(y) t}}\Big(\frac 12 - \frac{a(y)^2(x-y)^2}{4t}\Big)\,,
\end{split}
\]
by \eqref{paop0} and an integration by parts we get,
\[
\begin{split}
& \frac{\partial(\rr{e}^{Lt} v_\perp)(x)}{\partial x} = \int\! \frac{\rr{e}^{-\frac{a(y)(x-y)^2}{4 t}}}{\sqrt{4\pi a(y)t}} v_\perp'(y)\, \rr{d}y \\ & + \int\! \frac{a'(y) \rr{e}^{-\frac{a(y)(x-y)^2}{4 t}}}{a(y)\sqrt{4\pi a(y) t}}\Big(\frac{a(y)^2(x-y)^2}{4t} - \frac 12\Big) v_\perp(y)\, \rr{d}y \\ & + \int\! \frac{\partial Q_t(x,y)}{\partial x} v_\perp(y)\,\rr{d}y\,.
\end{split}
\]
By \eqref{paop}$_1$, for some positive constants $C_1$ and $\lambda$,
\[
\Big|\frac{\partial(\rr{e}^{Lt} v_\perp)(x)}{\partial x}\Big| \le \int\! \frac{C_1\rr{e}^{-\frac{(x-y)^2}{4\lambda t}}}{\sqrt{4\pi\lambda t}} (|v_\perp(y)|+|v_\perp'(y)|)\, \rr{d}y\,,
\]
which implies, recalling that the heat kernel is a contraction in $L^2$,
\[
\Big\|\frac{\partial(\rr{e}^{Lt} v_\perp)}{\partial x}\Big\|_{L^2} \le C_1\|(|v_\perp|+|v_\perp'|)\|_{L^2} \le 2C_1 \|v\|_{H^1}\,.
\]

If $t>1$ we write,
\[
\begin{split}
& \frac{\partial(\rr{e}^{Lt} v_\perp)(x)}{\partial x} = \int\! \frac{\partial\rr{e}^{L}(x,y)}{\partial x} (\rr{e}^{L(t-1)} v_\perp)(y)\,\rr{d}y
\end{split}
\]
and observe that, for any $f\in L^2(\bb R;\rr{d}x)$, by \eqref{paop0} and \eqref{paop}$_1$ for $t=1$, for some $C_2>0$,
\[
\Big\|\frac{\partial(\rr{e}^{L} f)}{\partial x}\Big\|_{L^2} \le C_2 \|f\|_{L^2}\,.
\]
Therefore, in view of \eqref{pao7bis}$_1$, setting $C_3=C_0C_2$,
\[
\begin{split}
\Big\|\frac{\partial(\rr{e}^{Lt} v_\perp)}{\partial x}\Big\|_{L^2} & \le C_2 \big\|\rr{e}^{L(t-1)} v_\perp \big\|_{L^2} \le C_3 \rr{e}^{-\alpha (t-1)} \|v_\perp\|_{L^2} \le C_3 \rr{e}^{-\alpha t} \|v_\perp\|_{H^1}\,.
\end{split}
\]
In conclusion, there is a constant $C_4>0$ such that 
\begin{equation}
\label{pao7tris}
\|\rr{e}^{Lt} v - \langle v, q \rangle_a q\|_{H^1} \le C_4 \rr{e}^{-\alpha t} \|v-\langle v, q \rangle_aq\|_{H^1}\,.
\end{equation}

\section{Non-linear stability}
\label{s:nonlineare}

In this section we prove the one-dimensional manifold $\mc M := \{\bar u_z\colon z\in \bb R\}$ of the translates of $\bar u$ is asymptotically stable. This is the content of two theorems. 

The first result follows from the linear analysis done in the previous section. Indeed, $\lambda=0$ is a semisimple eigenvalue of the (unbounded) operator $L$ on $H^1$, and the estimate \eqref{pao7tris} shows further that the remaining part of the spectrum is strictly contained in $\{\lambda\colon \Re{\lambda}<0\}$. This implies, see, e.g., \cite[Proposition 4.1]{MRS}, the following stability theorem. 

\begin{theorem}
\label{thm:1}
Consider the Cauchy problem 
\begin{equation}
\label{nle}
\begin{cases}
{\displaystyle \frac{\partial u}{\partial t} = \sigma(u) \Big[\frac{\partial^2u}{\partial x^2} -W'(u)\Big]}\,, \\ u(x,0) =u_0(x)\,.
\end{cases}
\end{equation}
For any $\varepsilon>0$ there exists $\delta>0$ such that if the initial datum $u_0$ satisfies $\|u_0-\bar u_{\zeta_0}\|_{H^1} \le \delta$ for some $\zeta_0\in \bb R$ then the solution $u(t)=u(\cdot, t)$ exists for all $t\ge 0$ and $\mathrm{dist}_{H^1}(u(t),\mc M) \le \varepsilon$ for any $t\ge 0$. Moreover, if $\delta$ is small enough, there is $z_0\in\bb R$ such that, for some $\beta>0$,
\[
\lim_{t\to +\infty} \rr{e}^{\beta t} \|u(t)-\bar u_{z_0}\|_{H^1} = 0
\]
and $|z_0-\zeta_0|\to 0$ if $\delta \to 0$.
\end{theorem}

In Theorem \ref{thm:2} below, we instead prove convergence in sup-norm to a suitable front of those solutions whose initial datum approximately resembles a front. The proof is an adaptation of the argument in the seminal paper \cite{FM}, where the case of constant mobility is concerned. In \cite{FM} it is also shown that the limit is reached exponentially fast in time. Due to the non-constant mobility, we are able to prove this rate of convergence only under the more restrictive assumption on the initial datum of Theorem \ref{thm:1}.

\begin{theorem}
\label{thm:2}
Consider the Cauchy problem Eq.~\eqref{nle} and assume that $u_0$ is a piecewise continuous function such that $0\le u_0\le 1$. Then, there exists a unique bounded classical solution $u(x,t)$ and $0 \le u(x,t) \le 1$. Furthermore, if
\begin{equation}
\label{bc}
\limsup_{x\to -\infty} u_0(x) < \bar r\,, \;\; \liminf_{x\to +\infty} u_0(x) > \bar r\,,
\end{equation}
with $\bar r$ as in \eqref{condW}, there is $z_0\in\bb R$ such that
\[
\lim_{t\to+ \infty} \| u(\cdot,t)-\bar u_{z_0} \|_\infty = 0\,.
\]
\end{theorem}

\noindent\textit{Proof.} We split the proof into three steps.

\noindent\textit{Step 1: Existence and uniqueness.} The existence and uniqueness in large of bounded classical solution $ u(x,t)$ and that $0 \le u(x,t) \le 1$ follow from a priori estimates and comparison theorems for quasilinear parabolic equations, see, e.g., \cite{LSU}.

\noindent\textit{Step 2: A priori estimates.} There are $z_1,z_2\in\bb R$ and $q_0, \mu>0$ such that
\begin{equation}
\label{ape}
\bar u_{z_1}(x) -q_0 \rr{e}^{-\mu t} \le u(x,t) \le \bar u_{z_2}(x) + q_0 \rr{e}^{-\mu t} \,.
\end{equation}
We prove the lower bound, the other can be obtained similarly. We fix $q_0>0$ such that 
\[
q_0<\bar r < 1-q_0 <  \liminf_{x\to +\infty} u_0(x)\,,
\]
so that there exists $z_*$ large enough for which  $\bar u_{z_*}(x) - q_0 \le u_0(x)$ for any $x\in \bb R$. We then look for $z_1 > z_*$ and $\mu>0$ such that, setting 
\[
q(t) = q_0 \rr{e}^{-\mu t}\,, \quad z(t) = z_1 + (z_*-z_1) \rr{e}^{-\mu t}\,,
\]
the function
\[
u_-(x,t) := \max\{0;\bar u_{z(t)}(x)-q(t)\}
\]
is a subsolution. This clearly proves the estimate since $u_-(x,t) \le u(x,t)$ by the Comparison Theorem and $\bar u_{z_1}(x) -q_0 \rr{e}^{-\mu t} \le u_-(x,t)$ by construction.

Since $u=0$ is a solution, $u_-$ is a subsolution if $u_->0$ implies
\[
N[u_-] := \frac{\partial u_-}{\partial t} - \sigma(u_-) \Big[\frac{\partial^2u_-}{\partial x^2} -W'(u_-)\Big] \le 0\,.
\]
Now, if $u_->0$, we have $\bar u_z > q$ and 
\[
N[u_-] = - \dot z \bar u_z' - \dot q + \sigma(\bar u_z - q) [W'(\bar u_z-q) - W'(\bar u_z)]\,,
\]
where we used that $\bar u_z'' = W'(\bar u_z)$ and omitted the explicit dependence on $x$ and $t$. Since $q_0<\bar r<1-q_0$, $W'(0) = W'(1) =0$, and $W''(0)$ and $W''(1)$ are positive, there is $C>0$ such that $W'(1-q)-W'(1) < - 2C q $ and $W'(0)- W'(q) \le -2C q$ for any $q\in [0,q_0]$. Therefore, by continuity, there exists $\delta<q_0$ such that $W'(u-q) - W'(u) \le -Cq$ for any $(u,q)\in [1-\delta,1] \times [0,q_0]$ and for any $0 \le q \le u \le \delta$. We conclude that whenever $u_z \in [1-\delta,1]$ or $q \le \bar u_z \le \delta$ we have,
\[
N[u_-] \le - \dot z \bar u_z' - \dot q - (\bar C/C) q \le (\mu -\bar C/C) q\,,
\]
with $\bar C$ as in Eq.~\eqref{csigma} and where we used that $\dot q = -\mu q$, $\dot z>0$, and $\bar u_z'>0$. On the other hand, from the properties of $\bar u$ and the smoothness of $W$, there is $C_5>0$ such that $\bar u'\ge C_5$ whenever $\delta \le \bar u \le 1-\delta$ and $W'(\bar u_z-q) - W'(\bar u_z)\le C_5 q$ for any $q\in [0,q_0]$. Therefore, if  $\delta \le \bar u_z \le 1-\delta$,
\[
\begin{split}
N[u_-] & \le - C_5\dot z - \dot q + \bar C C_5 q \le [C_5 \mu (z_*-z_1) + (\mu + \bar C C_5)q_0] \rr{e}^{-\mu t}\,,
\end{split}
\]
where we used the upper bound in  Eq.~\eqref{csigma} and that $\dot z = -\mu(z_*-z_1)\rr{e}^{-\mu t}$. Therefore, choosing $\mu=\bar C/C$ and $z_1 = z_* + q_0(\mu+C_5)/(\mu C_5)$ we obtain $N[u_-] \le 0$ whenever $u_->0$. 

An immediate corollary of the proof of Eq.~\eqref{ape} is the stability of fronts. Indeed, suppose that the initial datum satisfies also $|u_0(x)-\bar u_{z_0}(x)| \le \varepsilon$ for some $z_0\in\bb R$ and small $\varepsilon>0$. Then, it is possible to choose $q_0= O(\varepsilon)$ and $|z_*-z_0| = O(\varepsilon)$, so that also $|z_1-z_0| + |z_2-z_0| = O(\varepsilon)$. In conclusion, $|u(x,t)-\bar u_{z_0}(x)| = O(\varepsilon)$ independently of $x\in \bb R$ and $t\ge 0$. 

\noindent\textit{Step 3: Convergence to fronts.}  We now prove that there exists $z_0\in \bb R$ such that, uniformly in $x\in \bb R$,  
\begin{equation}
\label{convne}
\lim_{t\to+\infty}|u(x,t)-\bar u_{z_0}(x)| = 0\,.
\end{equation} 

We first observe that, for a suitable constant $C_6>0$,
\begin{equation}
\label{stimu1}
\begin{split}
|1-u(x,t)| & + \bigg|\frac{\partial u}{\partial t}(x,t) \bigg| + \bigg|\frac{\partial u}{\partial x}(x,t) \bigg| + \bigg|\frac{\partial^2 u}{\partial x^2}(x,t) \bigg| \le C_6\big(\rr{e}^{-\alpha_1x} + \rr{e}^{-\mu t}\big)\,, 
\end{split}
\end{equation}
for $x>0$, $t\ge 0$,
and
\begin{equation}
\label{stimu2}
\begin{split}
|u(x,t)| & + \bigg|\frac{\partial u}{\partial t}(x,t) \bigg|  + \bigg|\frac{\partial u}{\partial x}(x,t) \bigg|  + \bigg|\frac{\partial^2 u}{\partial x^2}(x,t) \bigg| \le C_6\big(\rr{e}^{\alpha_0x} + \rr{e}^{-\mu t}\big)\,,
\end{split}
\end{equation}
for $x<0$ and $t\ge 0$.
The estimates in Eqs.~\eqref{stimu1} and \eqref{stimu2} concerning the undifferentiated function $u$ come from Eqs.~\eqref{pao2} and \eqref{ape}. Since $W'(0) = W'(1)=0$, these estimates also imply that $|W'(u(x,t))| \le C\big(\rr{e}^{\alpha_0x} + \rr{e}^{-\mu t}\big)$  for some $C>0$. The bounds on the differentiated function then follow from standard Schauder estimates for quasilinear parabolic equations, see, e.g., \cite{LSU}.

Then, by arguing exactly as in \cite[Lemma 4.4]{FM}, we conclude that for each $\delta>0$ the orbit $\{u(\cdot, t)\colon t\ge\delta\}$ is relatively compact in $C^2(\bb R)$. 

To prove Eq.~\eqref{convne}, we let $v$ be the following truncation of $u$,
\[
v(x,t) = \begin{cases} u(x,t) & \text{ for } |x|\le t \,, \\ 0 & \text{ for } x\le - t - 1 \,, \\ 1 & \text{ for } x \ge t + 1\,, 
\end{cases}
\]
where the interpolation is done  smoothly in such a way that the estimates Eqs.~\eqref{stimu1} and \eqref{stimu2} hold also for $v$ (possibly, with a larger constant $C_6$).

We next introduce the Lyapunov function
\[
F(v) = \int_\bb R \bigg[\frac{1}{2} \bigg|\frac{\partial v}{\partial x}\bigg| ^2 +  W(v)\bigg]\,\rr{d}x\,.
\]
It is bounded uniformly in $t$ in view of Eqs.~\eqref{stimu1}, \eqref{stimu2}, and since  $W(0)=W(1) = 0$. Moreover, letting $f(t) := F(v(\cdot, t))$, after integration by parts it is easily shown that $f$ is differentiable and
\[
\dot f (t) = - \int_\bb R \bigg(\frac{\partial^2 v}{\partial x^2} - W'(v)\bigg) \frac{\partial v}{\partial t}\ \,\rr{d}x\,.
\]
From the definition of $v$ it follows that $\dot f(t) = Q(t) + R(t)$ with
\[
Q(t) = - \int_\bb R \sigma(v) \bigg(\frac{\partial^2 v}{\partial x^2} - W'(v)\bigg)^2 \,\rr{d}x \le 0
\]
and, for a some $C_7>0$, 
\[
\begin{split}
|R(t)| & = \bigg|\int_{t\le |x| \le t+1} \bigg(\frac{\partial^2 v}{\partial x^2} - W'(v)\bigg) \frac{\partial v}{\partial t} \,\rr{d}x\bigg| \\ & \le  C_7 \int_{- t -1}^{-t} \big(\rr{e}^{\alpha_0 x} + \rr{e}^{-\mu t}\big)^2\,\rr{d}x + C_7 \int_t^{t +1} \big(\rr{e}^{-\alpha_1x} + \rr{e}^{-\mu t}\big)^2\,\rr{d}x\,.
\end{split}
\]
Therefore $\displaystyle \lim_{t\to +\infty} R(t) = 0$ so that $\displaystyle \limsup_{t\to +\infty}\dot f(t) \le 0$. Hence, since $f(t)$ is bounded uniformly in time, there must exist a diverging sequence $\{t_n\}$ such that $\displaystyle \lim_{n\to \infty}\dot f(t_n) = 0$. Since $\displaystyle \lim_{t\to +\infty} R(t) = 0$, we finally obtain $\displaystyle \lim_{n\to \infty} Q(t_n) = 0$. 

Now, by compactness, we can extract a subsequence $\{t_{n_k}\}$ such that $u(\cdot, t_{n_k})$, and hence also $v(\cdot, t_{n_k})$, converges  in $C^2(\bb R)$ to some function $\bar v$. Since $\displaystyle \lim_{k\to \infty} Q(t_n) = 0$ we deduce that 
\[
\begin{split}
& \int_{|x|\le K} \sigma(v) \big(\bar v'' - W'(\bar v)\big)^2 \,\rr{d}x \\ & \quad = \lim_{k\to \infty} \int_{|x|\le K} \sigma(v) \bigg(\frac{\partial^2 v(\cdot, t_{n_k})}{\partial x^2} - W'(v(\cdot, t_{n_k}))\bigg)^2 \,\rr{d}x \le \lim_{k\to \infty} Q(t_n) = 0 
\end{split}
\]
for all $K>0$,
so that $\bar v'' - W'(\bar v)=0$. Since we also have $\bar v(-\infty) = 0$ and $\bar v(+\infty) = 1$, by the uniqueness of fronts we deduce $\bar v= \bar u_{z_0}$ for some $z_0\in \bb R$. 
To conclude the proof of Eq.~\eqref{convne}, we simply recall that we have already proved the stability of fronts, so that the convergence of $u(\cdot,t_{n_k})$ to $\bar u_{z_0}$ implies that of the whole trajectory $u(\cdot, t)$. 
\qed

\section{Numerical check of stability results}
\label{s:num}

In this section we discuss the result of some numerical 
simulations aiming to check the stability properties stated in 
Theorems~\ref{thm:1} and \ref{thm:2}.
The strategy will be the following: i) we consider the particular solution 
$\bar{u}\in\mathcal{M}$ defined in Eq.~\eqref{pao1} of the stationary equation 
Eq.~\eqref{ac040}
associated with the problem Eq.~\eqref{ac020};
ii) we consider a profile obtained by modifying locally $\bar{u}$ 
(type I)
and a profile obtained by adding an oscillatory term to $\bar{u}$ in such a 
way that the assumptions of Theorem~\ref{thm:2} are satisfied (type II);
iii) the problem Eq.~\eqref{ac020} is solved numerically via the
finite elements method and the $H^1$, $L^2$, and $L^\infty$ 
norms of the difference between the solution at time $t$ and $t-\Delta t$
are computed as a function of time. 


Our results will be plotted in figures made of two panels: in the left 
one we will show the solution of the problem Eq.~\eqref{ac020} at different 
values of time; in the right panel the 
norm ($H^1$, $L^2$, and $L^\infty$) 
of the difference between the solution at time $t$ and $t-\Delta t$.
Different cases will correspond to different choices 
of the initial profile. 

In this section the function $W$ is chosen as 
\begin{equation}
\label{num000}
W(u)
=
\frac{1}{2}u^2(1-u^2)
\,,
\end{equation}
so that the stationary solution Eq.~\eqref{pao1} is
\begin{equation}
\label{num010}
\bar{u}(x)=\frac{e^x}{1+e^x}
\,.
\end{equation}
The stiffness function will be 
\begin{equation}
\label{num020}
\sigma(u)=A+B\sin(2\pi\nu u)
\end{equation}
with $A,B,\nu\ge0$; note that for $\nu>1/2$ we shall 
choose $B<A$ so that the condition $\sigma(u)>0$  
is safely satisfied.
In the following picture we shall report results for the case 
$\nu=5/2$; other possible values of $\nu$ has been considered 
and similar results have been found. 

The type I initial condition will be a continuous function 
obtained by substituting the 
$\bar{u}$ profile in the interval $[a,b]$ with an arc of parabola 
with vertex in $c$, with $a<c<b$.
The type II initial condition will be a continuous function 
obtained by adding to  
$\bar{u}$ the function 
$D\sin[2\pi\mu(x-d)]$ 
in the interval $[d,\infty)$, with 
$0<D\le1/4$, $d\ge1$, and $\mu>0$.

Eq.~\eqref{ac020} is solved via the finite element method 
powered with the Newton--Raphson algorithm over a finite domain.
The boundary conditions are of Neumann type.
The spatial discretization of the domain is based on quadratic shape functions 
with one hundred elements. 
As time discretization we use the backward Euler scheme 
with time step $\Delta t =1$.
The algorithm is stopped when the $L^2$ norm of the difference 
between the current profile and the one computed one time step before 
is smaller than $10^{-4}$.

\begin{figure}
\centering
\hspace{-.4cm}
  \includegraphics[width=0.47\textwidth]{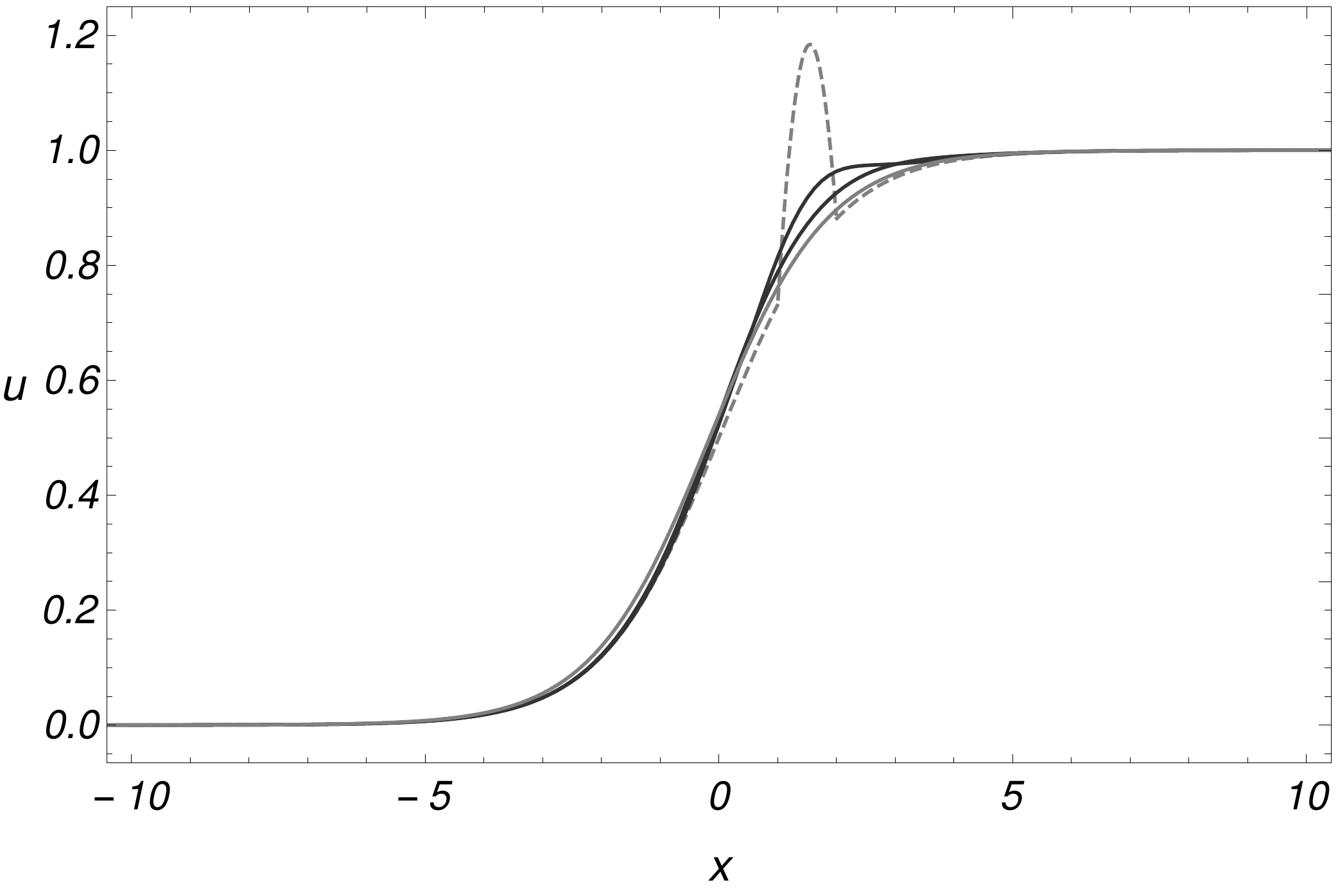}
\hspace{.1cm}
  \includegraphics[width=0.50\textwidth]{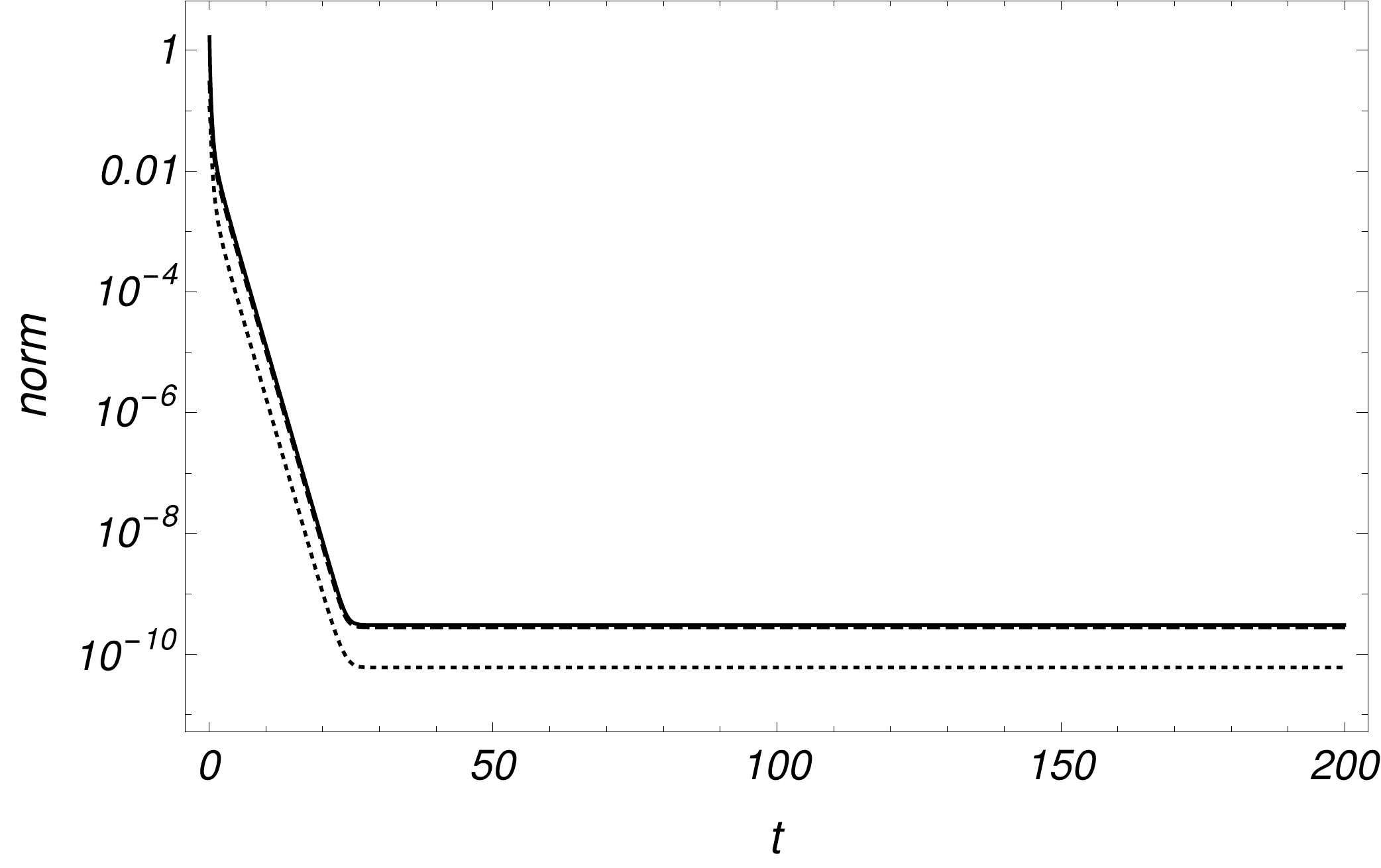}
\caption{Solution of problem Eq.~\eqref{ac020} for the initial condition 
of type I with $a=1$, $b=2$, and $c=1.55$ and stiffness Eq.~\eqref{num020}
with $A=1$, $B=1/2$, and $\nu=5/2$.  
Left panel: profile $u(x,t)$ at times $t=0$ (dashed gray), $5$ (solid black),
$10$ (solid black), and the stationary 
infinite time limit (solid gray).  
Right panel: semi--logarithmic plot of $H^1$ norm (solid line), 
$L^2$ norm (dashed line), 
$L^\infty$ norm (dotted line) 
of the difference between the solution at time $t$ and $t-\Delta t$.
}
\label{f:num020}
\end{figure}

\begin{figure}
\centering
\hspace{-.4cm}
  \includegraphics[width=0.47\textwidth]{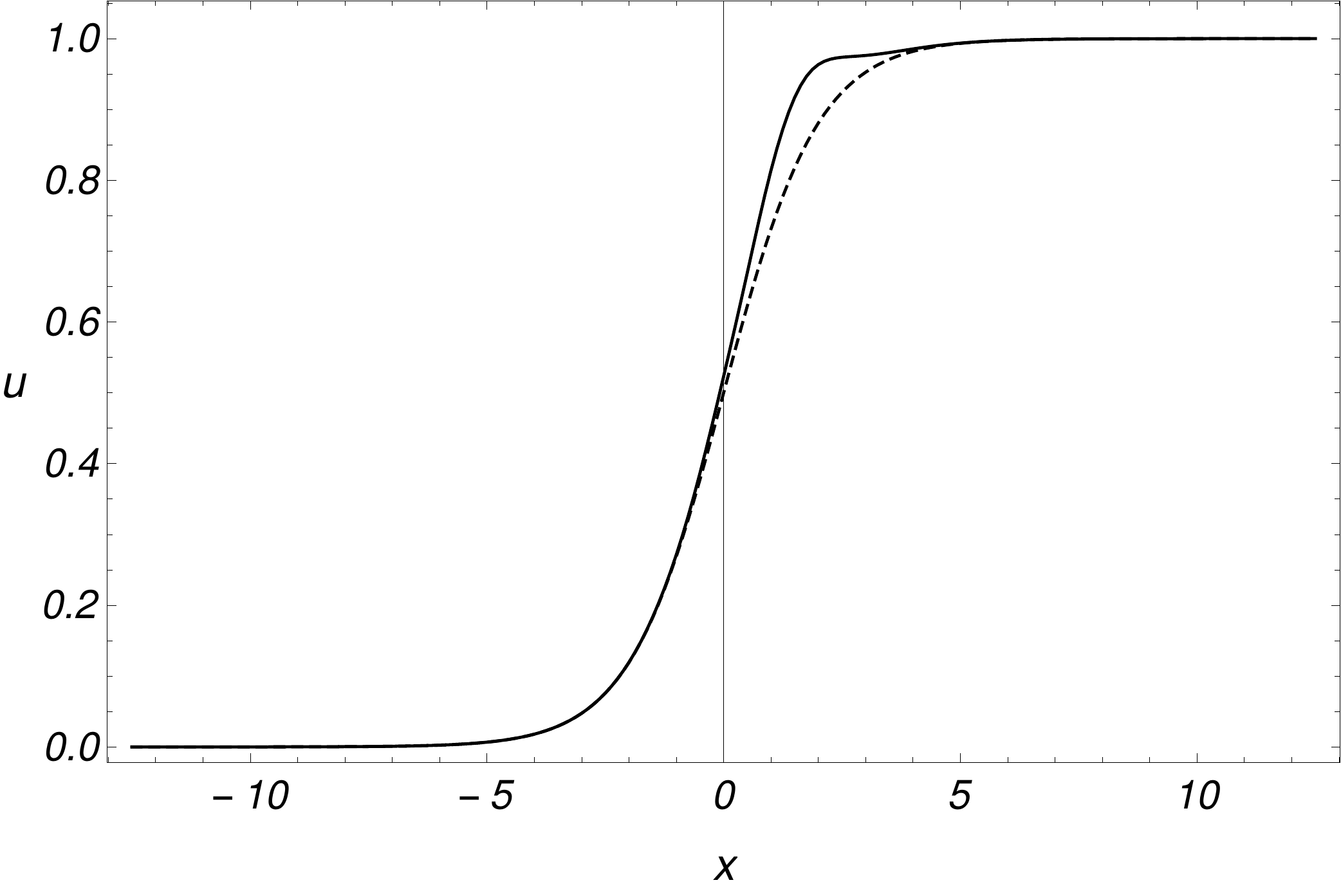}
\hspace{.1cm}
  \includegraphics[width=0.47\textwidth]{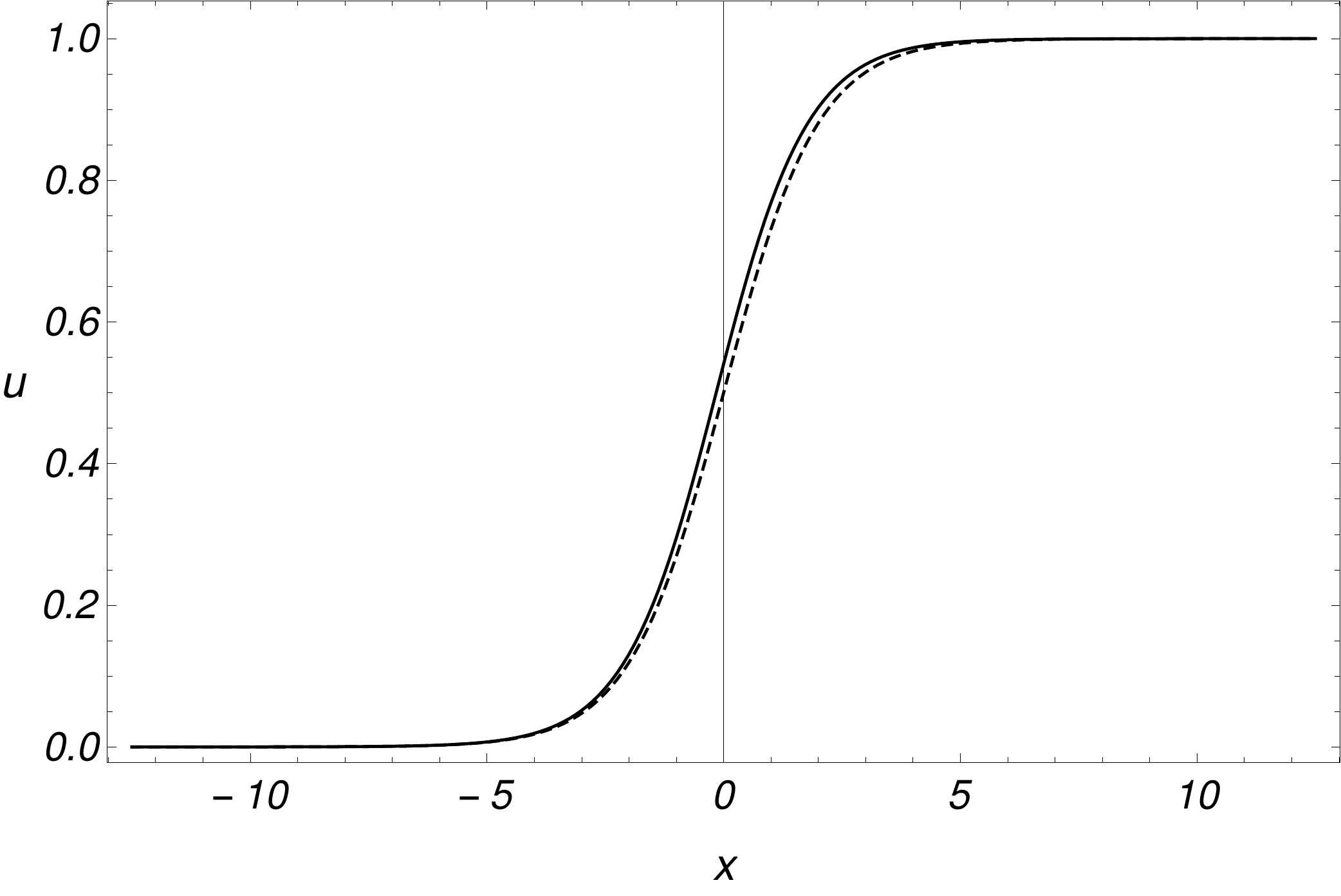}
\caption{Solid lines represent the profiles at times $t=5$ (left) 
and $t=25$ (right) of the evolution described in Figure~\ref{f:num020}.
The dashed line is the stationary profile already reported in 
Figure~\ref{f:num020}.}
\label{f:num030}
\end{figure}

In Figure~\ref{f:num020} an initial condition of type I is used (see the 
caption for more details) and, as already remarked, 
for the stiffness Eq.~\eqref{num020} we chose the frequency $\nu=5/2$. 
The solution of the equation converges to a solution 
in $\mathcal{M}$ slightly different from $\bar{u}$ which, we recall, 
has been perturbed to construct the initial condition. 
The behavior 
of the norm as a function of time is compatible with an exponential decay
and show three different regimes. An initial quick drop is followed 
by a linear decrease and, finally, a long constant behavior onsets. 
As shown in 
Figure~\ref{f:num030}, where we report the profile corresponding 
to the times where the steepness of the norm changes abruptly,
during the first part of the evolution 
the parabolic bump is removed, whereas in the second part 
the profile is slowly deformed to reach the stationary shape. 

\begin{figure}
\centering
\hspace{-.4cm}
  \includegraphics[width=0.47\textwidth]{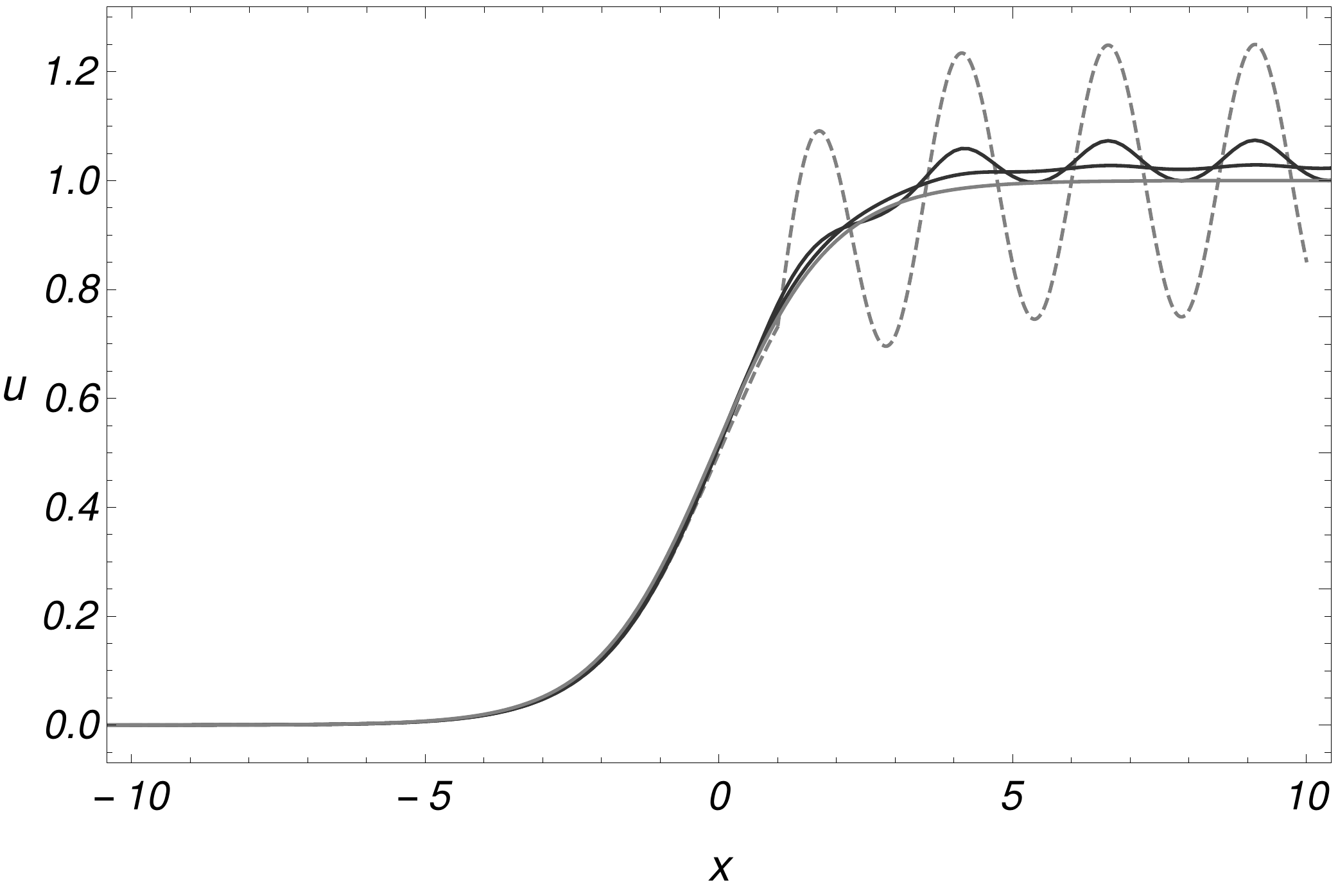}
\hspace{.1cm}
  \includegraphics[width=0.50\textwidth]{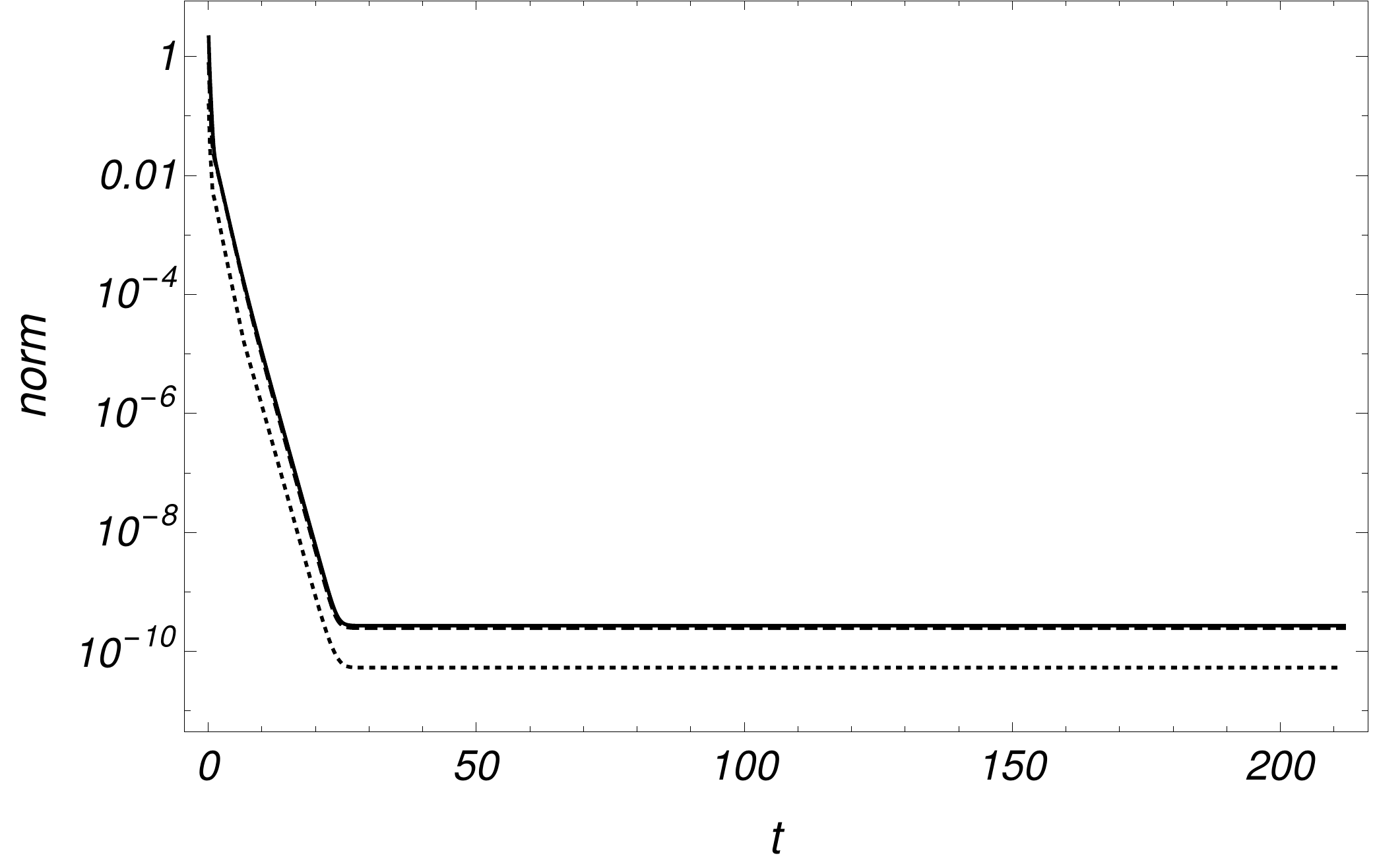}
\caption{Solution of problem~\eqref{ac020} for the initial condition 
of type II with $d=1$, $D=1/4$, and $\mu=0.4$ and stiffness Eq.~\eqref{num020}
with $A=1$, $B=1/2$, and $\nu=5/2$.  
Left panel: profile $u(x,t)$ at times $t=0$ (dashed gray), $5$ (solid black),
$10$ (solid black), and the stationary 
infinite time limit (solid gray).  
Right panel: semi--logarithmic plot of $H^1$ norm (solid line), 
$L^2$ norm (dashed line), 
$L^\infty$ norm (dotted line) 
of the difference between the solution at time $t$ and $t-\Delta t$.
}
\label{f:num040}
\end{figure}

\begin{figure}
\centering
\hspace{-.4cm}
  \includegraphics[width=0.47\textwidth]{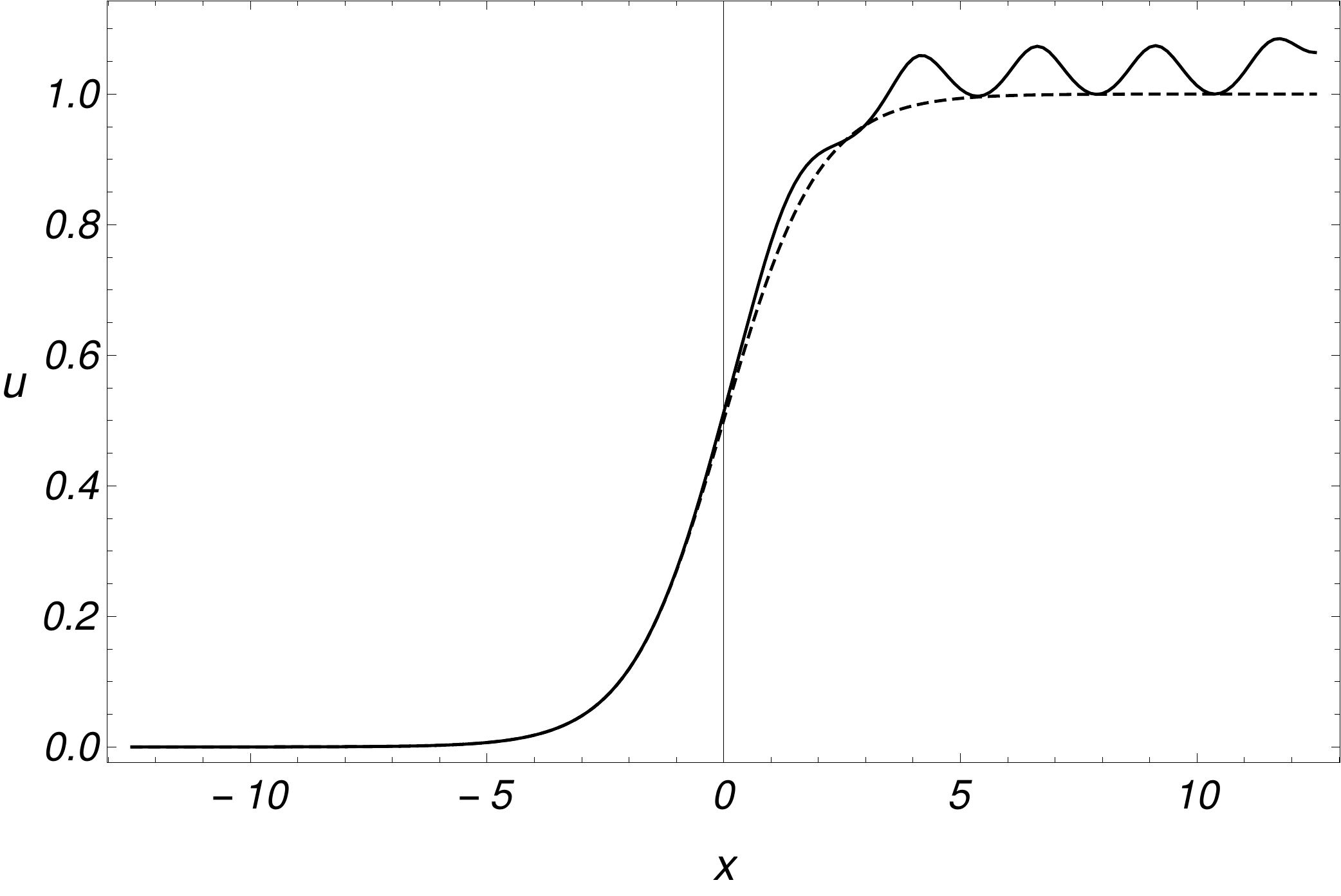}
\hspace{.1cm}
  \includegraphics[width=0.47\textwidth]{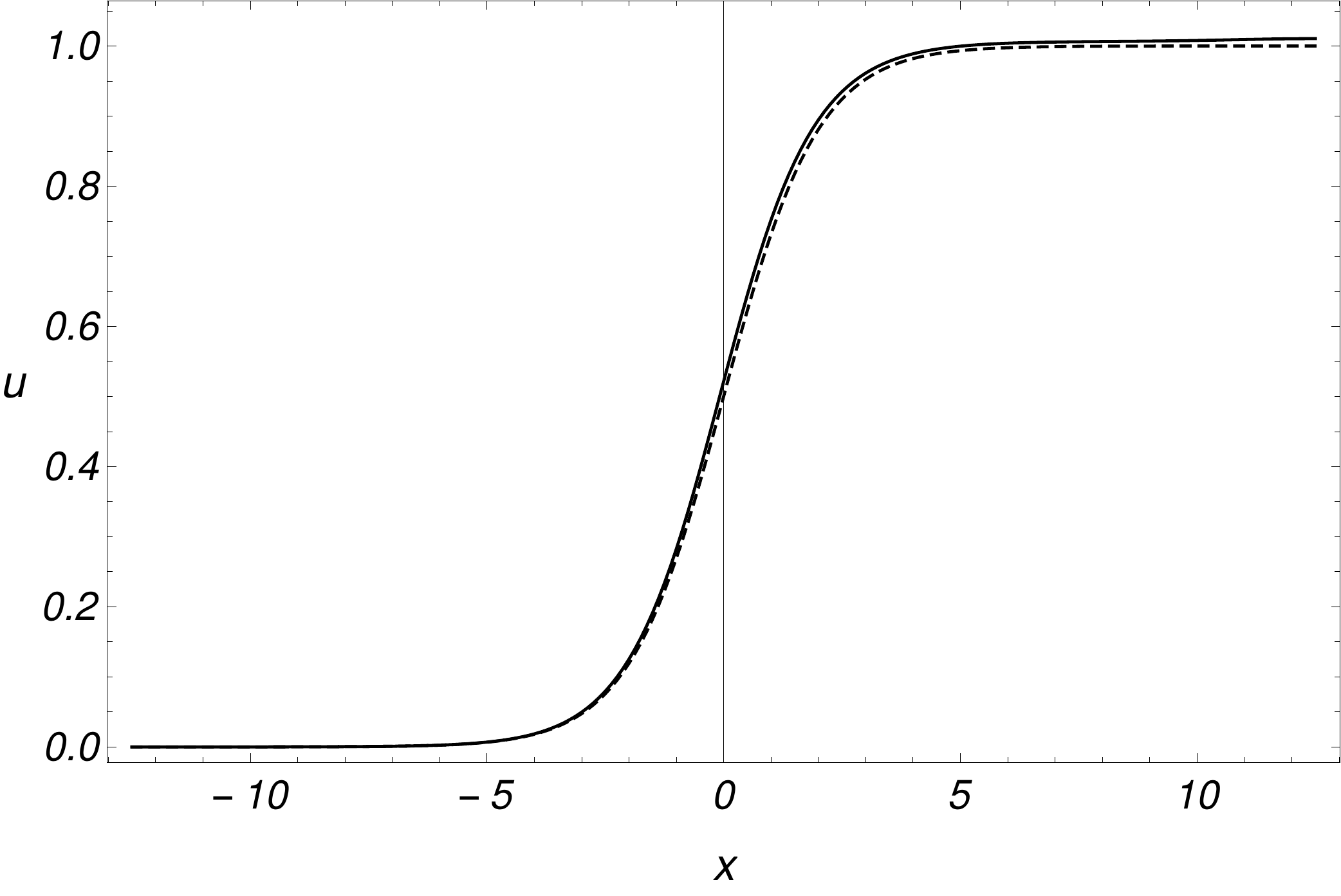}
\caption{Solid lines represent the profiles at times $t=5$ (left) 
and $t=25$ (right) of the evolution described in Figure~\ref{f:num040}.
The dashed line is the stationary profile already reported in 
Figure~\ref{f:num040}.}
\label{f:num050}
\end{figure}

In Figure~\ref{f:num040} an initial condition of type II is used (see the 
caption for more details) and 
for the stiffness Eq.~\eqref{num020} we chose, again, the frequency $\nu=5/2$.
The solution of the equation quickly converges to a monotone 
interface profile with constant plateau lower than one. 
Then, the solution starts to increase the value of the 
plateau slowly approaching the correct limiting value one. 
The behavior of the norm as a function 
of time is compatible with an exponential decay, also in this case. 
The three regimes already noted in Figure~\ref{f:num020} can be 
observed and explained as we did above. Indeed, as illustrated in 
Figure~\ref{f:num050} in the first part of the evolution, 
corresponding to the sudden drop of the norms, 
the ripples are removed and then the profile linearly 
converges to the stationary profile. 

\begin{figure}
\centering
\hspace{-.4cm}
  \includegraphics[width=0.47\textwidth]{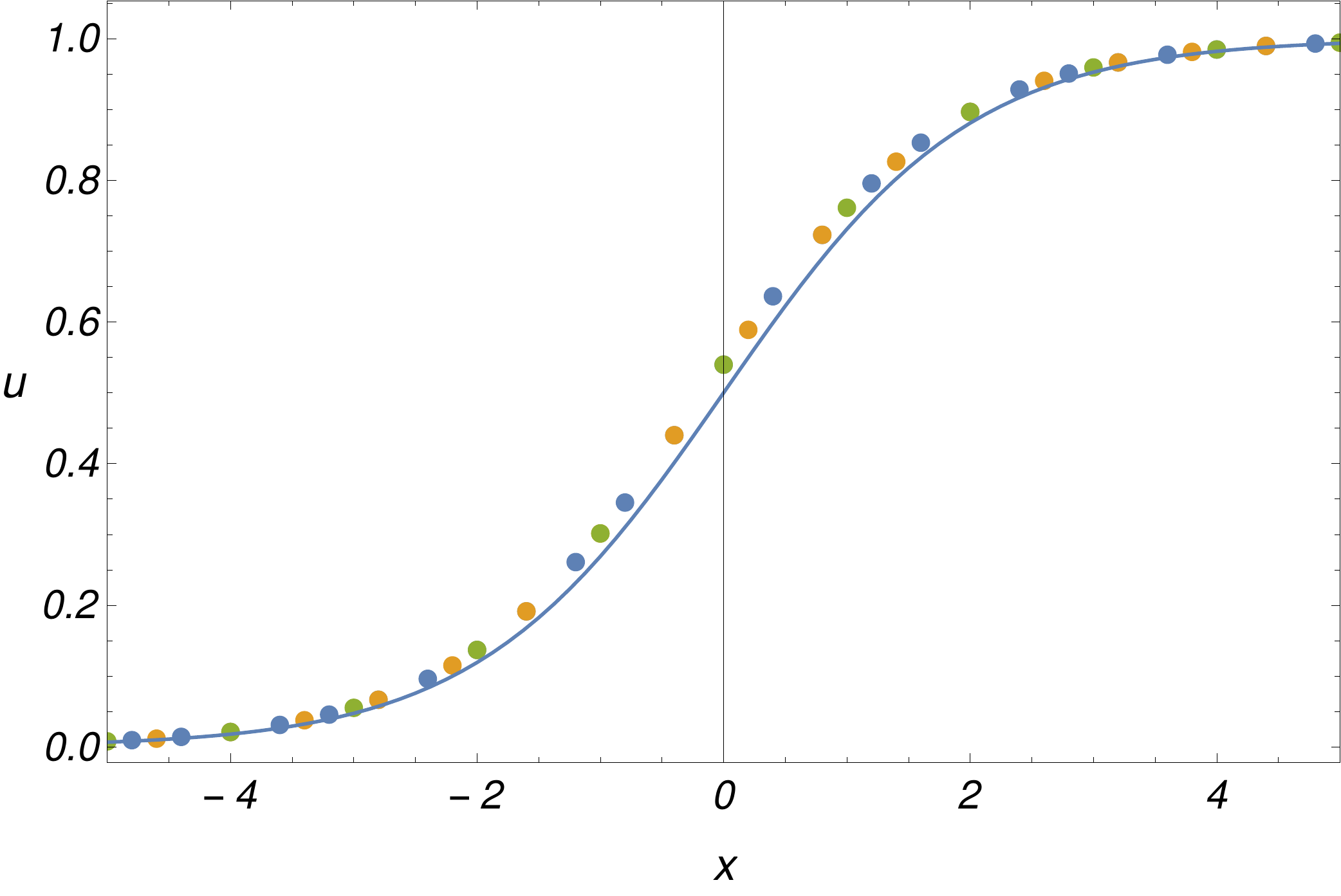}
\hspace{.1cm}
  \includegraphics[width=0.47\textwidth]{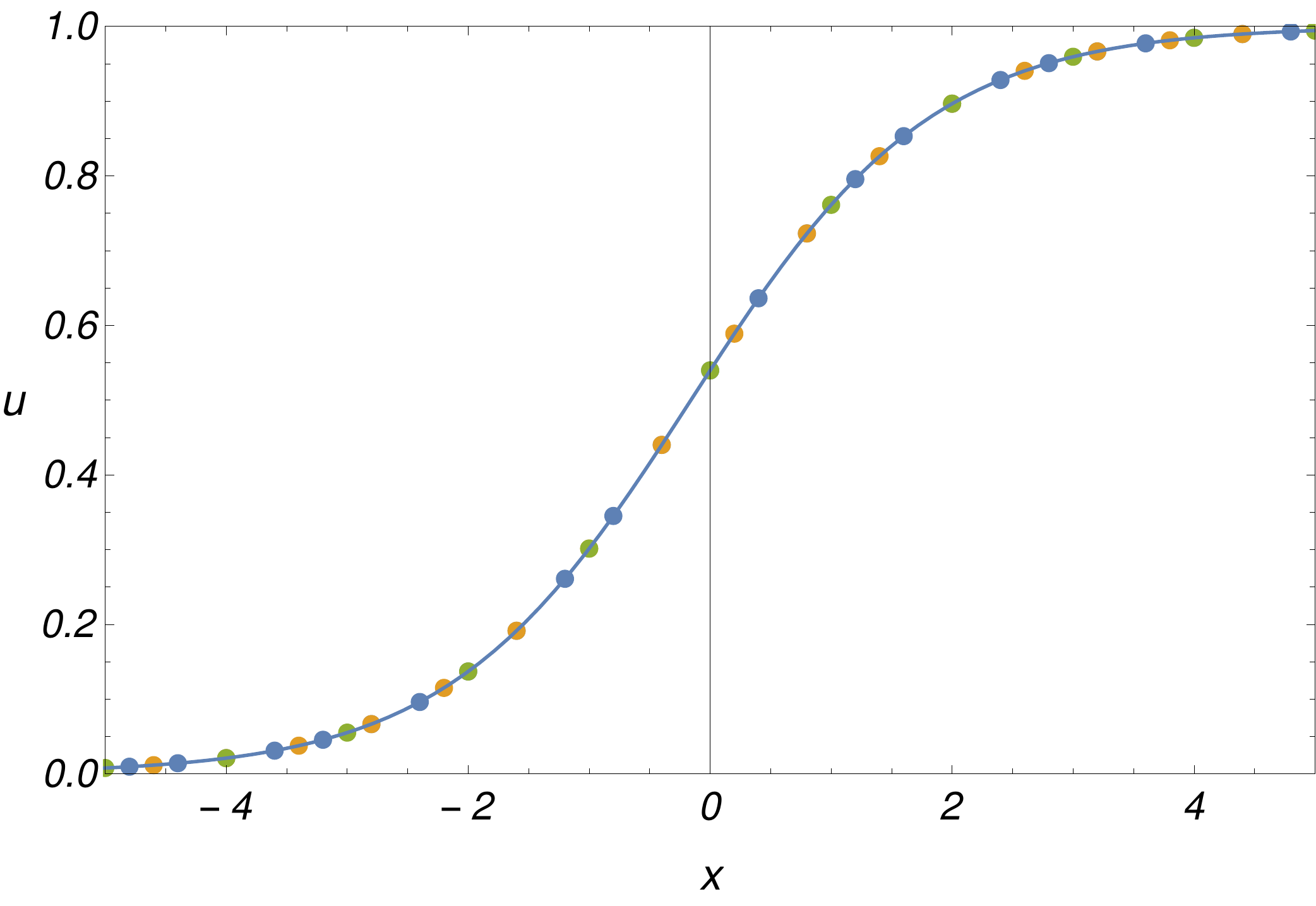}
\caption{On the left, exact solution \eqref{num010}
and
stationary profiles obtained for the simulation 
discussed in Figure~\ref{f:num020} on the intervals 
$[-10,10]$ (yellow dots), $[-12.5,12.5]$ (blue dots), and
$[-15,15]$ (green dots).
On the right: the same, but the exact solution has been shifted 
by the amount $0.15881$ for a perfect match.
}
\label{f:num060}
\end{figure}

Finally, we note that 
our simulations have been performed on a finite 
interval, whereas the theoretical discussion of the 
previous sections refer to the evolution on the real infinite line. 
In Figure~\ref{f:num060} we, thus, compare results 
obtained in simulations referring to 
different sizes of the space interval.
Since,
the stationary profiles obtained for the three different 
cases coincide (see the left panel),
we can infer that 
our simulations capture, indeed, the infinite volume 
behavior.
Moreover, we also note that by suitably shifting the 
exact profile, as shown in the right panel, 
we obtain a perfect match with the stationary profiles 
obtained on the finite space interval. 

\section*{Acknowledgments}
ENMC thanks D.\ Andreucci for some very useful discussions 
and for having provided some references.

\bibliography{bcs-allen01}

\end{document}